\documentclass[twocolumn,showpacs,preprintnumbers,amsmath,amssymb]{revtex4}

\usepackage{graphicx}
\usepackage{dcolumn}
\usepackage{bm}

\begin{document}


\title{Interacting Phantom Energy}

\author{Zong-Kuan Guo}
\email{guozk@itp.ac.cn}
\affiliation{%
Institute of Theoretical Physics, Chinese Academy of Sciences,
P.O. Box 2735, Beijing 100080, China
}%
\author{Yuan-Zhong Zhang}
\affiliation{
CCAST (World Lab.), P.O. Box 8730, Beijing 100080\\
Institute of Theoretical Physics, Chinese Academy of Sciences,
P.O. Box 2735, Beijing 100080, China
}%

\date{\today}

\begin{abstract}
We investigate the role of a suitable interaction between a matter
fluid and a phantom field for the coincidence problem.
There exists a stationary scaling solution which is a stable attractor
at late times. Furthermore, the cosmic doomsday is avoided in one
region of the parameter space.
\end{abstract}

\maketitle


Scalar field plays an important role in modern cosmology. Dark
energy can be attributed to the dynamics of a scalar field, called
quintessence~\cite{RP, ZWS}, which convincingly realize the present
accelerated expansion of the universe by using late-time attractor
solutions in which the
scalar field mimics the perfect fluid in a wide range of parameters.
But regarded as dark energy, quintessence field with the state equation
parameter $w > -1$ may be not consistent with recent
observation~\cite{ASSS}. In order to obtain $w < -1$, phantom field with
reverse sign in its dynamical term may be a simplest implementing
and can be regarded as one of interesting
possibilities describing dark energy~\cite{RRC}. The physical background
for phantom type of matter with strongly negative pressure would be
found in string theory~\cite{MGK}. Phantom field may also arise
from a bulk viscous stress due to particle production~\cite{JDB} or
in higher-order theories
of gravity~\cite{MDP}, Brans-Dicke and non-minimally coupled scalar
field theories~\cite{DFT}. The cosmological models which allow for
phantom matter appear naturally in the mirage cosmology of the
braneworld scenario~\cite{KK} and in k-essence models~\cite{COY}. In
spite of the fact that the field theory of phantom fields encounters the
problem of stability which one could try to bypass by assuming them
to be effective fields~\cite{GWG}, it is nevertheless interesting to
study their cosmological implication. Recently, there are many relevant
studies of phantom energy~\cite{SW}.

The physical properties of phantom energy are rather weird, as they
include violation of the dominant-energy condition, naive superluminal
sound speed and increasing energy density with time. The latter property
ultimately leads to unwanted future singularity called big rip. This
singularity is characterized by the divergence of the scale factor in a
finite time in future~\cite{CKW}. To avoid the cosmic doomsday, specific
scalar field models were proposed~\cite{CHT}. It requires a special class
of phantom field potentials with local maximum. Moreover, the energy
density of the phantom field increases with time, while the energy density
of the matter fluid decreases as the universe evolves. Why the energy
densities of dark matter and phantom energy are of the same order
just at the present epoch? In this paper we investigate the role of a
possible coupling of dark matter and phantom field for the coincidence
problem. With the help of a suitable coupling~\cite{ZPC, CJPZ}, we find
that there exists a stationary scaling solution and demonstrate numerically
that it is a stable attractor at late times. Furthermore, the cosmic
doomsday is avoided in one region of the parameter space.


We consider the case in which both the phantom energy with constant
$w < -1$ and the cold dark matter are present. Then, if the
universe ceases to be matter-dominated at cosmological time $t_m$,
the scale factor can be written as
\begin{equation}
a(t)=a(t_m)\left[-w+(1+w)\left(\frac{t}{t_m}\right)\right]^{2/3(1+w)}
\end{equation}
at $t > t_m$. It is easy to see that the scale factor blows up at
$t=wt_m/(1+w)$. This occurs because, even though the energy densities
in ordinary types of matter are redshifting away, the energy density of
phantom energy increases in an expanding universe. It is possible that
both components decrease with time if there is a transfer of energy from
the phantom field to the matter fluid. Then the cosmic doomsday may
be avoided.

For a spatially flat FRW universe with the matter fluid $\rho_m$ and
the phantom field $\phi$, Friedmann equation can be written as
\begin{equation}
\label{FE}
H^2=\frac{\kappa ^2}{3}\left(\rho_p+\rho_m\right),
\end{equation}
where $\kappa^2 \equiv 8\pi G_N$ is the gravitational coupling and
the energy density and pressure, $\rho_p$ and $P_p$, of the
homogeneous phantom field $\phi$ are given by
\begin{eqnarray}
\rho_p &=& -\frac{1}{2}\dot{\phi}^2+V(\phi), \\
P_p   &=& -\frac{1}{2}\dot{\phi}^2-V(\phi),
\end{eqnarray}
respectively, in which $V(\phi)$ is the phantom field potential.
We postulate that the two components, $\rho_p$ and $\rho_m$,
interact through the interaction term $Q$ according to
\begin{eqnarray}
\label{EM1}
\dot{\rho}_p+3H\gamma_p \, \rho_p &=& -Q, \\
\dot{\rho}_m+3H\gamma_m\rho_m &=& Q,
\label{EM2}
\end{eqnarray}
where
\begin{eqnarray}
\gamma_p & \equiv & 1+w = \frac{\rho_p+P_p}{\rho_p} \\
\gamma_m & \equiv & \frac{\rho_m+P_m}{\rho_m}
\end{eqnarray}
satisfy $\gamma_p \le 0$ and $1 \le \gamma_m \le 2$. The interaction
term $Q$ represents an additional degree of freedom which can be
specified by the existence of solution with a stationary energy density
ratio $r \equiv \rho_m / \rho_p$ at late times. Using Eqs.(\ref{EM1})
and (\ref{EM2}) we obtain the evolution equation of the ratio $r$:
\begin{equation}
\label{RE}
\dot{r}=r\left(\frac{Q}{\rho_m}+\frac{Q}{\rho_p}
 -3H\gamma_m+3H\gamma_p\right).
\end{equation}
Obviously, the suitable interaction
\begin{equation}
\label{QE}
Q=3H(\gamma_m-\gamma_p)\frac{r\rho_p}{1+r}
\end{equation}
guarantees the existence of the stationary solution. Note that $Q > 0$,
which implies there is a transfer of energy from the phantom field to
the matter fluid.

We assume an interaction characterized by $Q=3Hc^2 (\rho_p+\rho_m)$
where $c^2$ denotes the transfer strength, which has already been
discussed in Ref.~\cite{ZPC}. When $\gamma_p$ and $\gamma_m$
are assumed to be constants, the two stationary solutions to Eq.(\ref{QE})
are
\begin{equation}
r_s^\pm=\frac{\gamma_m-\gamma_p}{2c^2}-1 \pm
 \sqrt{\left(\frac{\gamma_m-\gamma_p}{2c^2}-1\right)^2-1} \, ,
\end{equation}
which imply $0 < r_s^- \le 1 \le r_s^+$ and $r_s^- r_s^+=1$ when
$\gamma_m-\gamma_p \ge 4c^2$. A stability analysis of the stationary
solution indicates that the matter-dominated scaling solution $r_s^+$
is unstable, while the phantom-dominated scaling solution
$r_s^-$ is stable~\cite{CJPZ}. The energy density ratio $r$ evolves from
the unstable stationary value $r_s^+ > 1$ to the stable stationary solution
$r_s^- < 1$, which is clearly compatible with the presently favored
observational data $\Omega_m \approx 0.3$ and
$\Omega_\Lambda \approx 0.7$.
This may provide us with the dynamics of the density ratio that is
relevant to the solution of the coincidence problem.


Now consider the case of $\gamma_m=1$ (i.e. cold dark matter).
We will investigate the cosmological evolution with the stationary density
ratio $r$.
Substituting Eq.(\ref{QE}) into Eq.(\ref{EM1}), it follows that
\begin{equation}
\dot{\rho}_p+3\frac{\gamma_p+r}{1+r} H \rho_p = 0,
\end{equation}
which yields
\begin{equation}
\label{RP}
\rho_p \propto a^{-3\nu},
\end{equation}
where $\nu \equiv \frac{\gamma_p+r}{1+r}$. If the expansion is matter
dominated until the time $t_m$, then we can write the scale factor as
\begin{equation}
a(t)=a(t_m)\left(1-\nu+\nu \frac{t}{t_m}\right)^{2/3\nu}.
\end{equation}
Using the definitions of $r$ and $\gamma_p$, the Friedmann equation
(\ref{FE}) gives
\begin{equation}
a \propto
 \exp{\left(\frac{\sqrt{1+r}}{\sqrt{-3(1+w)}} \kappa \phi\right)}.
\end{equation}
In terms of $\phi$, (\ref{RP}) becomes
\begin{equation}
\label{RPE}
\rho_p \propto
 \exp{\left(-\frac{\sqrt{3(1+r)}\, \nu}{\sqrt{-(1+w)}}\kappa \phi\right)}.
\end{equation}
Since $\gamma_p$ is a constant, that is, potential and kinetic energies
of the phantom field remain proportional, it follows that
\begin{equation}
\label{VE}
V(\phi)=V_0
 \exp{\left(-\frac{\sqrt{3(1+r)}\, \nu}{\sqrt{-(1+w)}}\kappa \phi\right)}.
\end{equation}
It is reassuring to find an exponential potential because this type
potentials arise very naturally in the models of unification, such
as Kaluza-Klein theories, supergravity theories and string theories.
Similar result has been obtained in Refs.~\cite{RP, CW}. These
authors started with an exponential potential in which there is a free
parameter and then investigated the parameter range for which there
exists a stable scaling solution. We have first constructed a solution
with expected properties and then derived the corresponding potential.
In the coupled phantom scenario, it is surprising that the signs of the
exponent in Eq.(\ref{VE}) are different in different regions of the
parameter space, which are determined by the signs of the parameter
$\nu$. We will find that the increasing potential corresponds to a
climbing-up phantom field while the decreasing potential corresponds
to a rolling-down phantom field.

The recent observations indicate that approximately 0.3 and 0.7 of
the total energy density of the universe attribute to dark matter
and dark energy, respectively. To solve the coincidence problem
of our present universe, the stable stationary ratio of the
energy densities should satisfy $r_s^- < 3/7$, which corresponds to
$c^2 < 0.21(1-\gamma_p)$ (i.e. the regions I and II in Figure 1).
In the region I
\begin{displaymath}
1-\frac{1}{1-\gamma_p} < c^2 < \frac{21}{100}(1-\gamma_p)
\end{displaymath}
for $-3/7 < \gamma_p < 0$, since $0 < \nu < 2/3$, the energy
density of the phantom field decreases in an expanding universe and
the universe accelerates without a big rip in the future. However, in
the region II
\begin{displaymath}
c^2 < 1-\frac{1}{1-\gamma_p}
\end{displaymath}
for $-3/7 < \gamma_p < 0$ and
\begin{displaymath}
c^2 < \frac{21}{100}(1-\gamma_p)
\end{displaymath}
for $\gamma_p < -3/7$, $\nu < 0$ indicates that the universe
accelerates until a big rip occurs at $t=(1-1/\nu)t_m$. We see that
the energy transfer from the phantom field to the cold dark matter
prolongs the lifetime of the universe.

\begin{figure}
\begin{center}
\includegraphics[scale=1]{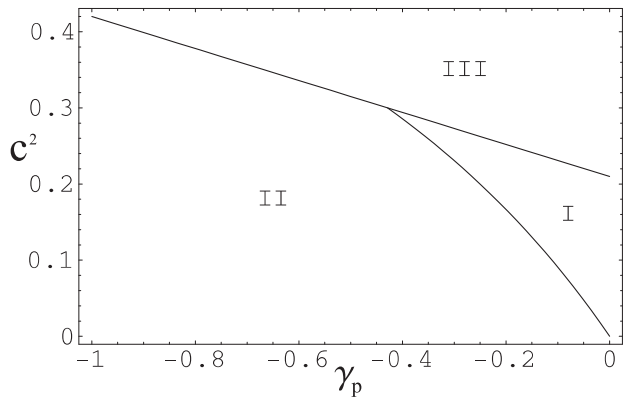}
\caption{Regions of the ($\gamma_p$, $c^2$) parameter space with
$\gamma_m=1$. In the regions I and II, the scaling
solution is a stable attractor. In the region III, there exist no stable
scaling solutions with $r_s^- < 3/7$. The universe accelerates for ever
in the region I, while the universe leads to a big rip in the future in
the region II.}
\end{center}
\end{figure}


To study an explicit numerical evolution of the energy density ratio $r$
and the phantom energy density (or the matter energy density), it is most
convenient to rewrite the evolution equations (\ref{RE}) and (\ref{EM1})
(or Eq.(\ref{EM2})) as a set of two first-order differential equations with
two independent variables $r$ and $\rho_p$ (or $\rho_m$)
\begin{eqnarray}
r' &=& 3c^2r
 \left(2+r+\frac{1}{r}-\frac{\gamma_m-\gamma_p}{c^2}\right), \\
\rho_p' &=& -3\rho_p \left(\gamma_p+c^2r+c^2\right), \\
\rho_m' &=& -3\rho_m \left(\gamma_m-c^2-\frac{c^2}{r}\right),
\end{eqnarray}
where the prime denotes a derivative with respect to the logarithm of the
scalar factor, $N \equiv \ln a$. Let us consider two points, ($-0.2$, $0.2$)
and ($-1.4$, $0.2$) in the regions I and II of the ($\gamma_p$, $c^2$)
parameter space in Figure 1, respectively. We choose different initial
conditions and follow the evolution. We find that an initial energy density
ratio decreases and there exist an attractor curve, which corresponds to
the stationary scaling solution. It implies that the universe evolves from
a dark matter dominance to a phantom dominance and the energy
density ratio of the two components becomes a constant ultimately.
Furthermore, we find that the stationary energy density ratio increases as
the transfer strength $c^2$ increases when $\gamma_p$ is fixed.
In Figures 2 and 3, the energy densities of
the two components redshift away. From Eqs.(\ref{RPE}) and (\ref{VE}),
we note that the phantom field rolls down the exponent potential.
In Figures 4 and 5, the energy density of
the phantom field increases with time, while the energy density of the
cold dark matter decreases initially and then increases ultimately with time.
From Eqs.(\ref{RPE}) and (\ref{VE}), we note that the phantom field
climbs up the exponent potential.


We have considered above that the present accelerated expansion
of our flat FRW universe is driven by an interacting mixture of a
matter fluid and a phantom field with $w < -1$. In the absence of
interaction, as shown in Ref.~\cite{GPZZ}, there exist no scaling
solutions because the phantom energy increases while the matter energy
decreases with time.
With the help of a suitable coupling, there exists a stable, stationary
scaling solution, which requires a transfer of energy from the phantom
field to the matter fluid.
Furthermore, we demonstrate numerically that an interaction between
cold dark matter and phantom field can drive the transition from
a matter dominance to a phase of accelerated expansion with a
stationary ratio of the energy densities of the two components.
This interacting phantom approach indicates a phenomenological
solution of the coincidence problem.
The different regions in the ($\gamma_p$, $c^2$) parameter space lead
to different fates of the universe. In the region I,
the phantom field rolls down the exponent potential and the universe
accelerates without the cosmic doomsday. However, in the region II, the
phantom energy increases as the phantom field climbs up the potential,
which lead to the divergence of scale factor in the future.
Coupling of the quintessence field to dark matter may be worrisome
because of quantum corrections to the quintessence potential~\cite{DJ}.
It is valuable to study the stability of the phantom potential under
quantum fluctuations in the case with a coupling to dark matter.

{\bf Note added.} While this letter is under review, Ref.~\cite{CRGW}
presents a similar idea.

\section*{Acknowledgements}
It is a pleasure to acknowledge helpful discussions with Winfried Zimdahl.
This project was in part supported by National Basic Research Program
of China under Grant No.2003CB716300 and also by NNSFC under
Grant No.10175070.

\begin{figure}
\begin{center}
\includegraphics[angle=-90,scale=0.37]{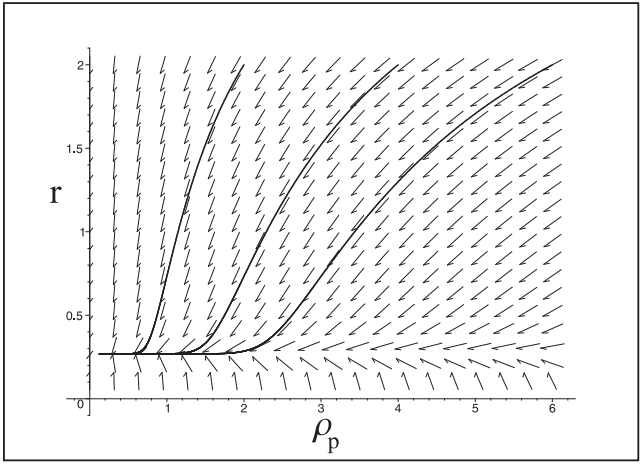}
\caption{$r$-$\rho_p$ with $\gamma_m=1$, $\gamma_p=-0.2$ and
$c^2=0.2$}
\end{center}
\end{figure}
\begin{figure}
\begin{center}
\includegraphics[angle=-90,scale=0.37]{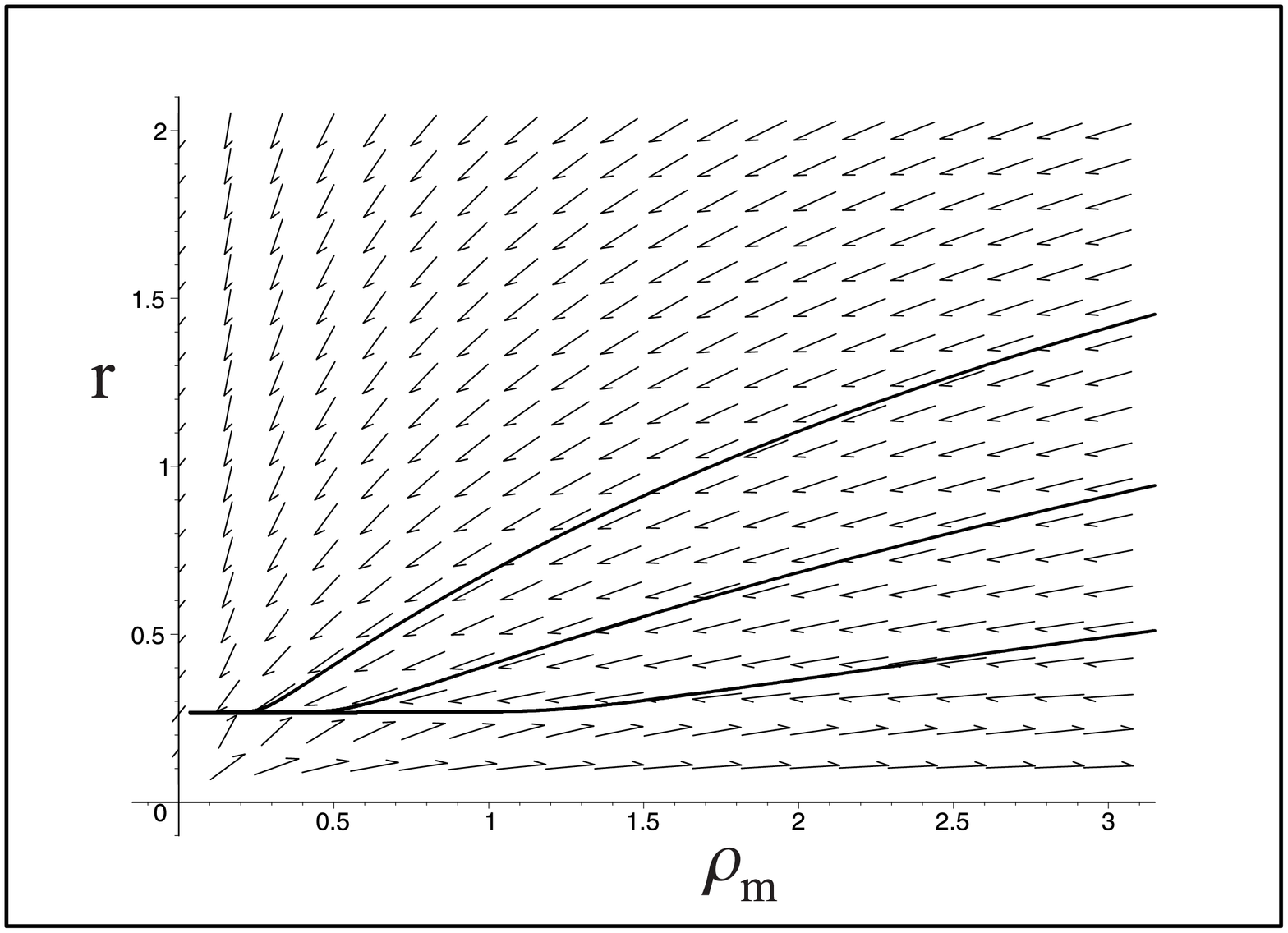}
\caption{$r$-$\rho_m$ with $\gamma_m=1$, $\gamma_p=-0.2$ and
$c^2=0.2$}
\end{center}
\end{figure}
\begin{figure}
\begin{center}
\includegraphics[angle=-90,scale=0.37]{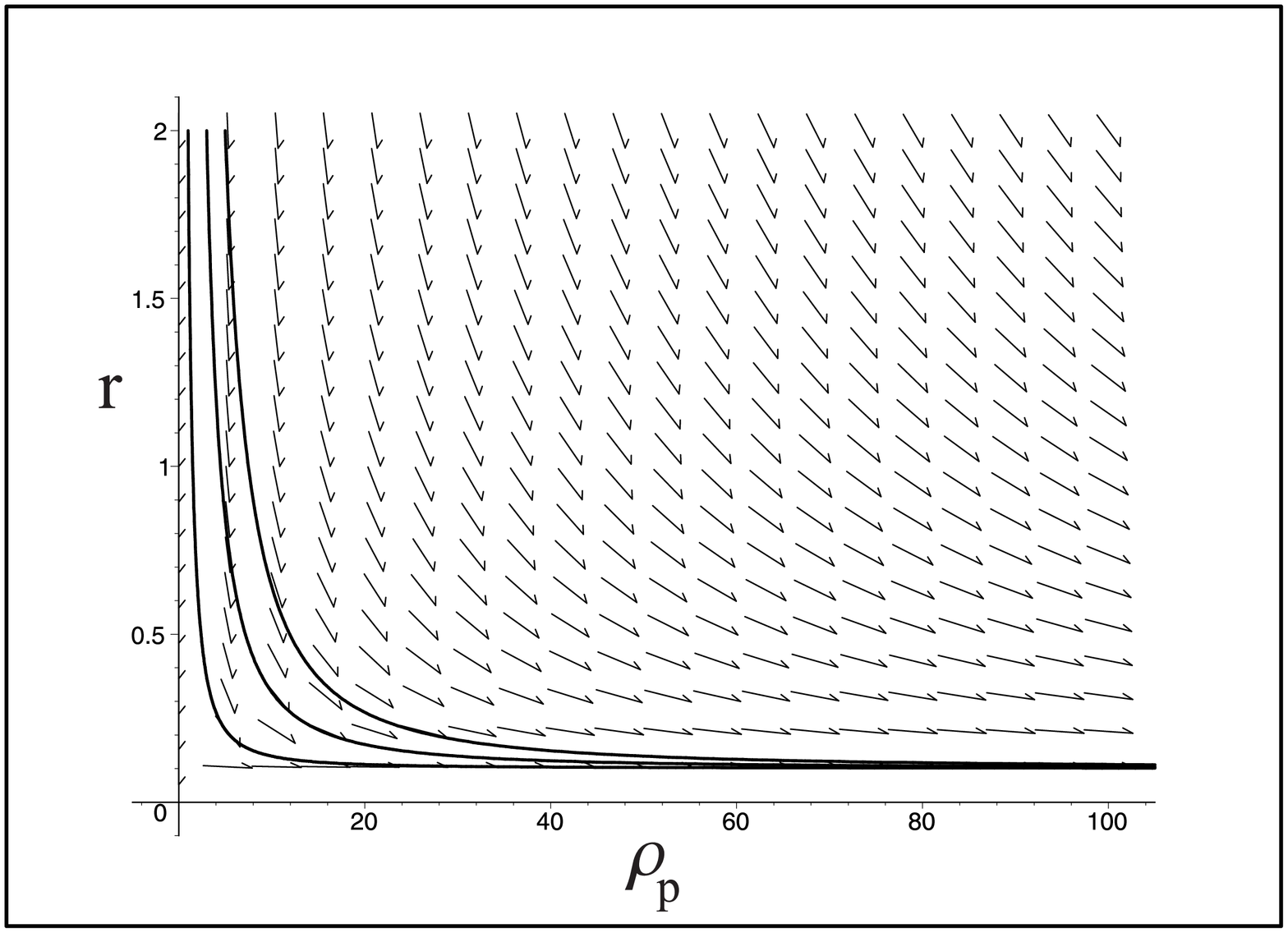}
\caption{$r$-$\rho_p$ with $\gamma_m=1$, $\gamma_p=-1.4$ and
$c^2=0.2$}
\end{center}
\end{figure}
\begin{figure}
\begin{center}
\includegraphics[angle=-90,scale=0.37]{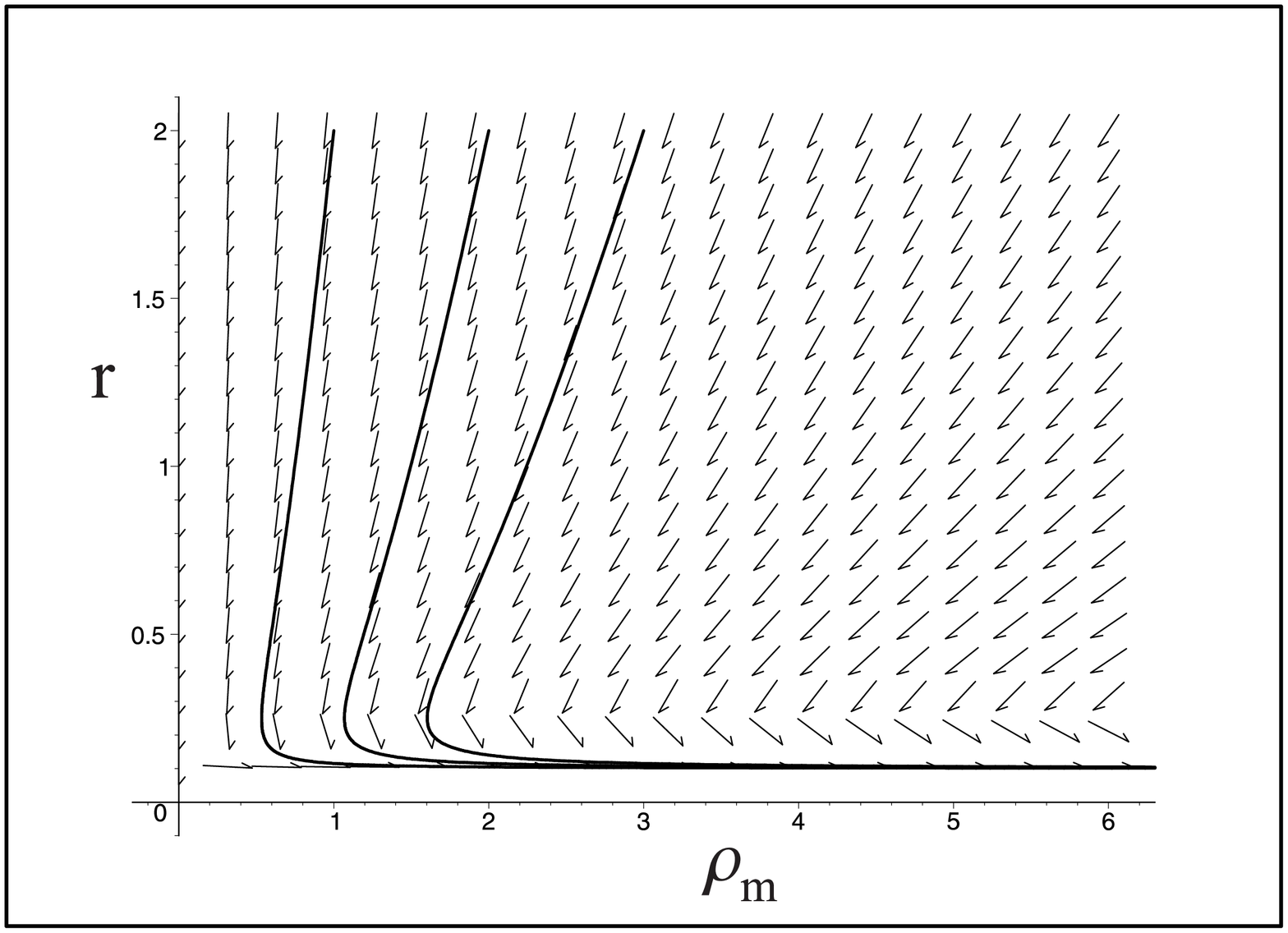}
\caption{$r$-$\rho_m$ with $\gamma_m=1$, $\gamma_p=-1.4$ and
$c^2=0.2$}
\end{center}
\end{figure}

\end{document}